\documentclass[11pt]{article}
\usepackage{amsmath,epsf,amssymb,latexsym,enumerate,cite,shadow,array,mcite,units,verbatim,amsmath}
\usepackage{slashed}
\usepackage{graphicx,wasysym}

\DeclareFontFamily{U}{rsf}{} \DeclareFontShape{U}{rsf}{m}{n}{
  <5> <6> rsfs5 <7> <8> <9> rsfs7 <10-> rsfs10}{}
\DeclareMathAlphabet\Scr{U}{rsf}{m}{n} \makeatletter
\@addtoreset{equation}{section} \makeatother

\def\be{\begin{equation}}
\def\ee{\end{equation}}
\def\ba{\begin{array}}
\def\ea{\end{array}}

\newcommand{\bea}{\begin{eqnarray}}
\newcommand{\eea}{\end{eqnarray}}

\def\K{K{\"a}hler}
\def\bi{{\bar I}}

\usepackage{ifpdf}
\ifx\pdfoutput\undefined
   \pdffalse
\else
   \pdfoutput=1
   \pdftrue
  \usepackage[pdftex]{hyperref}
\def\u0{{\underline 0}}

\def\url{{\underline {r+\ell}}}

\newcommand{\rf}[1]{(\ref{#1})}

\usepackage{color}
\usepackage[makeroom]{cancel}
\usepackage{graphicx}
\usepackage{amssymb,amsthm}
\usepackage{hyperref}

\def\bi{\begin{itemize}}
\def\ei{\end{itemize}}
\def\be{\begin{equation}}
\def\ee{\end{equation}}

\begin{document}

\begin{titlepage}

\hskip 1cm

\vskip 2.5cm

\begin{center}

{\LARGE \bf{Does the first chaotic inflation model in supergravity}} \\ \vskip 0.6 cm {\LARGE \bf  provide the best fit to the Planck data?}

\

\

\

{\bf  Andrei Linde} 
\vskip 0.2cm
{\small\sl\noindent SITP and Department of Physics, Stanford University, Stanford, CA
94305 USA}
\end{center}
\vskip 2.5 cm

\begin{abstract}
I describe the first model of chaotic inflation in supergravity, which was proposed by Goncharov and the present author in 1983. The inflaton potential of this model has a plateau-type behavior $V_{0} (1- {8\over 3}\, e^{-\sqrt 6 |\phi|})$ at large values of the inflaton field. This model predicts $n_{s} = 1-{2\over N} \approx 0.967$ and $r = {4\over 3 N^{2}} \approx 4 \times 10^{{-4}}$, in good agreement with the Planck data. I propose a slight generalization of this model, which allows to describe not only inflation but also dark energy and supersymmetry breaking.

\end{abstract}

\vspace{24pt}
\end{titlepage}

\section{Introduction}

The chaotic inflation scenario  \cite{Linde:1983gd} was proposed in 1983
 as an alternative to new inflation,  when it was realized that the idea of high temperature phase transitions and supercooling, the trademark of old and new inflation, made inflation very difficult to implement.  The simplest chaotic inflation models presented in   \cite{Linde:1983gd} were  models of the type $\phi^{n}$, but it was emphasized there that this scenario is much more general. The main idea of chaotic inflation was to consider various sufficiently flat potentials, either large-field or small-field, and check whether inflation may occur in some parts of the universe under some generic (and possibly chaotic) initial conditions, without making an assumption that the universe was in a state of thermal equilibrium and that initial state of the inflaton field should correspond to an extremum of the potential. At present, this idea may seem nearly trivial;  all presently available inflationary models are based on it. However, this scenario, which does not even require the existence of the hot Big Bang, is so much different from the old cosmological paradigm that for several years since the invention of chaotic inflation many found it psychologically difficult to accept.  Even now most of the college textbooks continue describing inflation as an intermediate stage of expansion of the universe in a supercooled vacuum state formed after the hot Big Bang. 
 

In October 1983, Alexander Goncharov and I developed the first version of chaotic inflation in supergravity  \cite{Goncharov:1983mw}, which I will call GL model hereafter. It was quite economical, involving only a single chiral superfield. It was the first supergravity model with the inflaton potential asymptotically approaching a plateau $V   \sim a- b e^{-c\phi}$. Later on, it was realized that the Starobinsky model \cite{Starobinsky:1980te}, after certain generalizations \cite{Starobinsky:1983zz,Kofman:1985aw}, can be cast in a form with a similar plateau potential \cite{Whitt:1984pd}. It took almost 30 years until the models of this type attracted general attention because they were strongly favored by the recent WMAP and Planck data \cite{Hinshaw:2012aka,Ade:2013uln}. Paradoxically, at that time the GL model was nearly forgotten.


 In this paper I will briefly revisit the GL model. I will describe its predictions in terms of $n_{s}$ and $r$, compare it with other related models, and then, following \cite{Ferrara:2014kva,Kallosh:2014via,Dall'Agata:2014oka,Kallosh:2014hxa}, I will propose a slight generalization of this model, which allows to describe not only inflation, but also the present stage of acceleration with a tiny cosmological constant $\sim 10^{{-120}}$ and with a controllable level of supersymmetry breaking.

 \section{The GL model}
The model proposed in  \cite{Goncharov:1983mw} described a single chiral superfield $\Phi$ with the simple \K\ potential $\Phi\bar\Phi$ and a superpotential $e^{-\Phi^{2}/2}\, W(\Phi)$, which was designed  to cancel the growth of the potential $V(\Phi)$ in the direction of the real part of the field $\Phi = {1\over\sqrt{2}}(\phi+i\chi)$. Using \K\ invariance, one can represent the GL model  \cite{Goncharov:1983mw} in an equivalent but simpler form as a theory with a shift-symmetric \K\ potential
\be\label{shift}
K = -{1\over 2} (\Phi-\bar\Phi)^{2}
\ee
and the superpotential
\be\label{s}
W = {m\over 6} \sinh{\sqrt{3}\Phi}\, \tanh{\sqrt{3}\Phi} \ .
\ee
 The superpotential (2.2) can be also written in a more symmetric form,  
\be\label{cosh}
W = {m\over 6}\, \bigl(\cosh{\sqrt{3}\Phi} - \cosh^{-1}{\sqrt{3}\Phi}\bigr)\ .
\ee

\begin{figure}[ht!]
\begin{center}
\includegraphics[width=10cm]{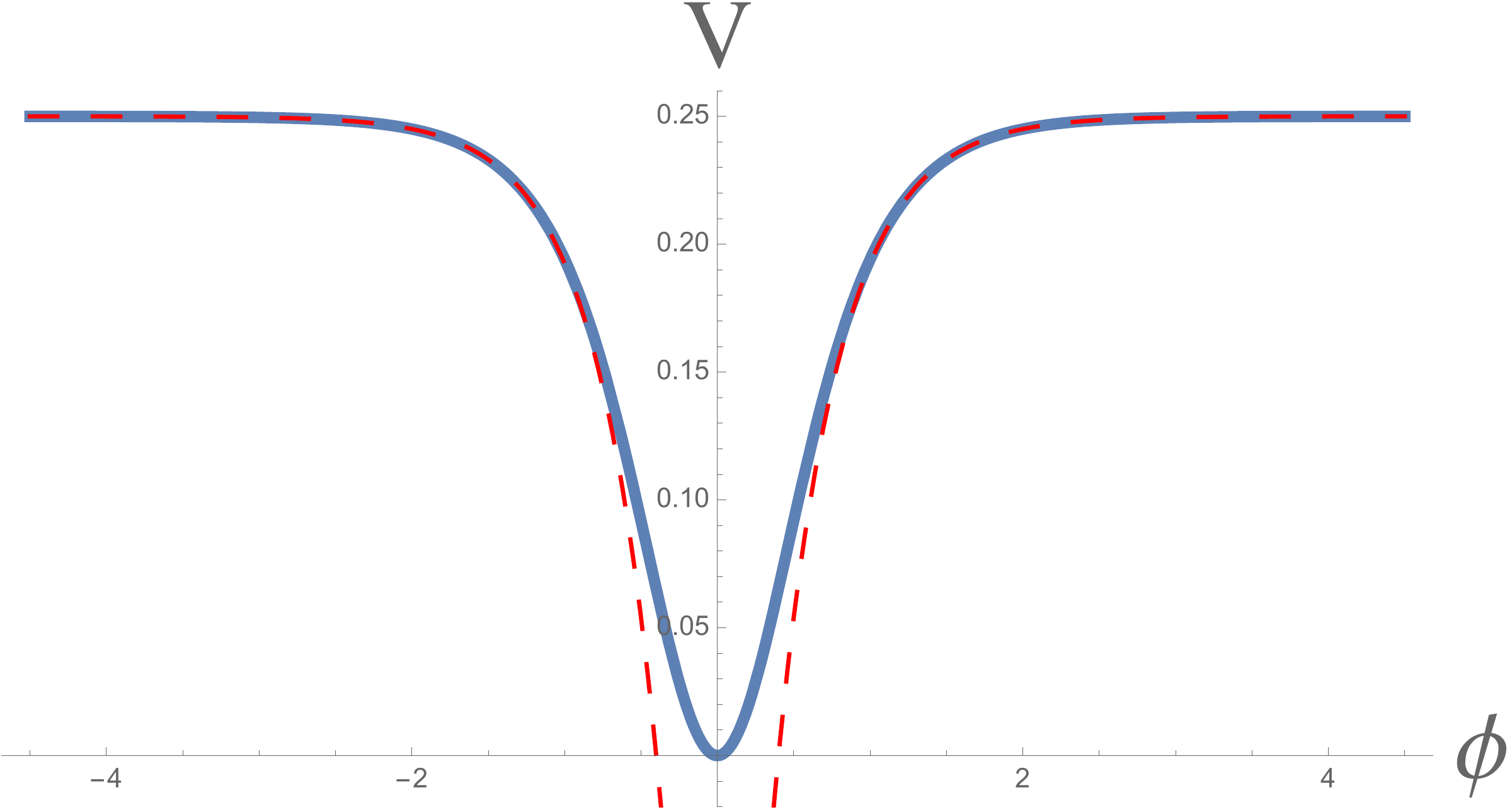}
\caption{\footnotesize The thick blue line shows the inflaton potential  \rf{pot} in the theory \rf{shift}, (2.2)  in units $m = 1$. The red dashed line shows its asymptotic representation \rf{app}, which exponentially rapidly converges to  in the inflationary regime with $\phi \gtrsim 1$. The last 60 e-foldings of the evolution of the universe correspond to $\phi \lesssim 2.8$. }\label{2}
\end{center}
\end{figure}

The potential $V(\phi,\chi)$ of the fields $\phi$ and $\chi$ in this model has a minimum at $\phi = \chi = 0$, where it vanishes, $V(0)  =0$. Because of the shift symmetry of the \K\ potential in the $\phi$, direction, the potential does not grow as $e^{\phi^{2}/2}$ in the $\phi$ direction, but blows up as $e^{\chi^{2}}$ in the $\chi$ direction.  As a result, it has a deep flat valley in the $\phi$ direction, with a minimum at $\chi = 0$. The mass squared of the field $\chi$ during inflation is very large, $m^{2}_{\chi} \gg H^{2}$. Therefore the field $\chi$ vanishes and plays no role during inflation.
Meanwhile the potential of the field $\phi$, which plays the role of a canonically normalized inflaton field, is given by
\be\label{pot}
V(\phi)= {m^2\over 12}  \Bigl(4 -  \tanh^{ 2}\sqrt{3\over 2} \phi\Bigr)\,  \tanh^{ 2}\sqrt{3\over 2} \phi\ \ .
\ee
This potential has a minimum at $\phi = 0$, where it vanishes,  see Fig. \ref{2}. At $\phi \gtrsim 1$, the potential coincides with 
\be\label{app}
V(\phi) = {m^{2}\over 4} \left(1- {8\over 3}\, e^{-\sqrt 6 |\phi|}\right) \ ,
\ee 
up to exponentially small higher order corrections  $O(e^{-3\sqrt 6 |\phi|})$  \cite{Goncharov:1983mw}. These corrections can only lead to higher order corrections in $1/N$ to $n_{s}$ and $r$, where $N \sim 60$ is the number of e-foldings. 

A proper interpretation of this model can be given in the context of the recently discovered theory of superconformal $\alpha$-attractors   \cite{Ferrara:2013rsa,Kallosh:2013yoa}. 
It will be shown in \cite{KL} that the GL superpotential (2.2) is the simplest superpotential in the family of superconformal attractors with a single chiral superfield. For the theory of two fields with the superpotential $Sf(\Phi)$, the general form of the $\alpha$-attractor superpotential is $S f(\tanh{\Phi\over \sqrt {3 \alpha}})$  \cite{Kallosh:2013yoa}, whereas for the single-field superpotential, the general form is $\sinh \sqrt  3\, \Phi\  f(\tanh \sqrt  3\,\Phi)$ \cite{KL}. The potential \rf{pot}, \rf{app} is of the same type as the potentials in the $\alpha$-attractor models  \cite{Kallosh:2013yoa} for $\alpha = 1/9$; in this respect see also \cite{Ellis:2013nxa}. 

Investigation of the slow-roll regime in this model is quite simple: One can use the well known equation relating to each other the field $\phi$ and the number of e-foldings $N$, 
\be
{d\phi\over dN} = {V'\over V} \approx {8\sqrt 6\over 3} e^{-\sqrt 6 |\phi|}
\ee
and find that $\phi = {1\over \sqrt 6} \log (16 N)$.
This means, for example, that the structure of the universe on a scale corresponding to $N = 60$ e-foldings was formed when the inflaton field was 
$\phi_{60} \sim 2.8$, and for $N = 50$ one finds $\phi_{50} \sim 2.7$. Thus one may argue that the observational predictions of this theory are mostly determined by the potential $V(\phi)$ in a small field interval close to $\phi = 2.8$.

The mass of the inflaton field at the minimum of the potential can be calculated using the Planck normalization of the amplitude of scalar perturbations of metric. It is given by $m  \sim 7 \times 10^{-6}$.
The most important observational prediction of this theory is its prediction for $n_{s} $ and $r$. In limit of  large number of e-foldings $N$, one has
\be\label{alpha}
n_{s} = 1-{2\over N} \approx 0.967\ , \qquad r = {4\over 3 N^{2}} \approx 3.7 \times 10^{-4} \ .
\ee
where the numerical values are given for $N = 60$. The first of these two relations was presented in  \cite{Goncharov:1983mw} not in terms of $n_{s}$, but in a more precise way as a logarithmic dependence of the amplitude of perturbations on the wavelength. The numerical values change just a bit for $N \sim 55$: $n_{s}  \sim 0.963$, $r \sim 4.4 \times 10^{-4}$.

These results are in a very good agreement with the observational data by WMAP \cite{Hinshaw:2012aka} and Planck 2013 
\cite{Ade:2013uln}. The prediction of the GL model for $n_{s}$ coincides with the prediction of the Starobinsky model   \cite{Starobinsky:1980te}, of the Higgs inflation \cite{Salopek:1988qh} and of a broad set of the cosmological attractor models discovered during the last two years \cite{Kallosh:2013yoa,Kallosh:2013hoa,Kallosh:2013daa}. The prediction for $r$ coincides with the prediction of $\alpha$-attractors for $\alpha = 1/9$ \cite{Kallosh:2013yoa}; it is 9 times smaller than the predictions of the Starobinsky model and of the Higgs inflation for $r$.

The preliminary results of the  Planck 2014 data release indicate that the models of this type may provide the best fit to the new set of observational data \cite{Finelli},  but of course we should wait for the final Planck 2014 data release, which should include a combined analysis of the observational data by Planck and BICEP2.

\section{Generalized GL models as cosmological attractors}
\begin{figure}[htb]
\begin{center}
\includegraphics[width=9cm]{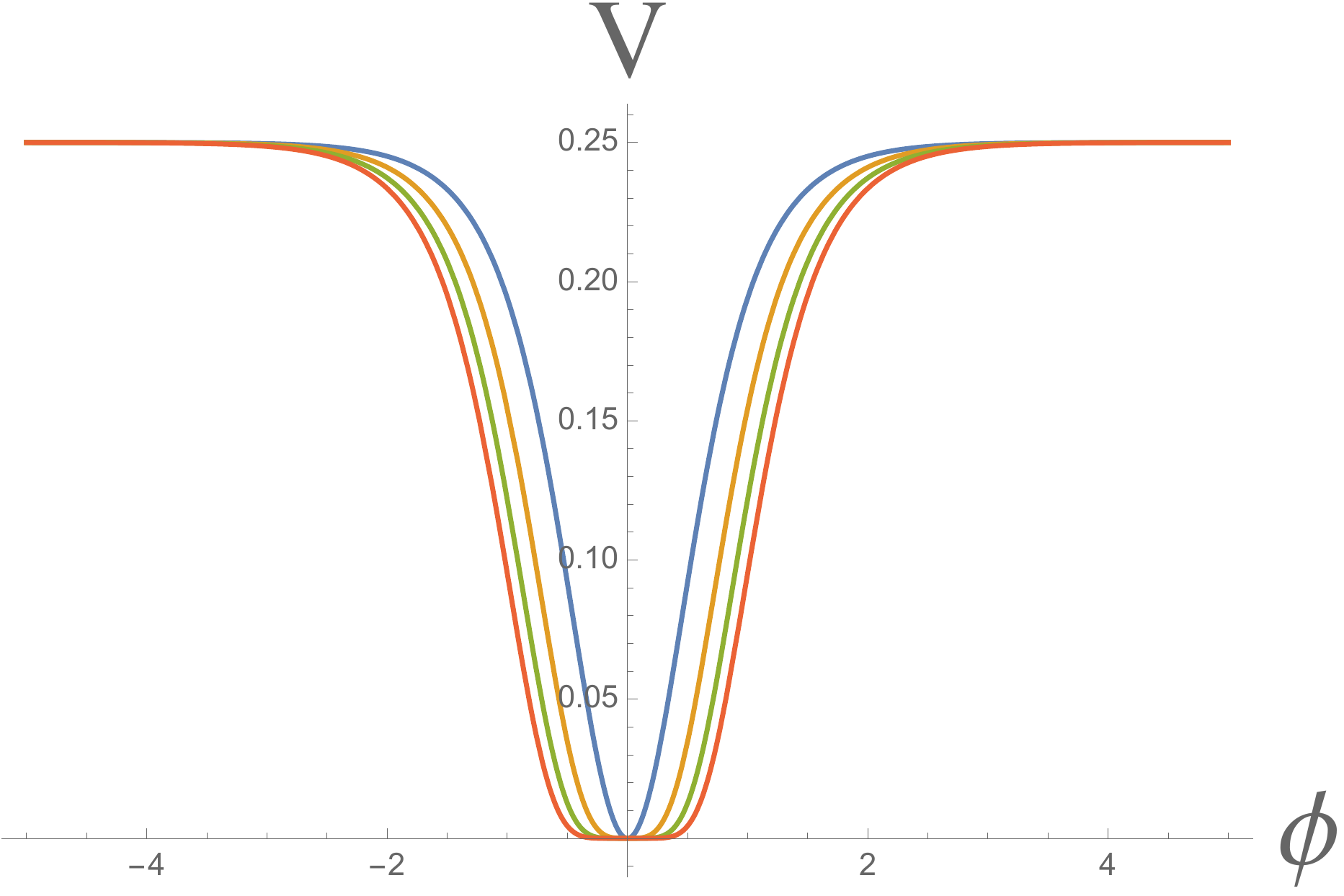}
\caption{\footnotesize The scalar potential in the generalized GL model \rf{sn}. The potentials for $n = 1$, 2, 3, 4 are shown by blue, yellow, green and red lines correspondingly.}\label{3}
\end{center}
\vspace{-0.5cm}
\end{figure}

The original GL model allows some generalizations. For example, one can consider superpotentials
\be\label{sn}
W = {m\over 2\sqrt{3+6 n} }\sinh{\sqrt{3}\Phi}\, \tanh^{n}{\sqrt{3}\Phi} \ ,
\ee
which lead to potentials shown in Fig. \ref{3} for $n = 1$, 2, 3, 4. This figure looks very similar to Fig.~1 in Ref. \cite{Kallosh:2013hoa}, which shows T-models belonging to the class of conformal attractors. At large $\phi$, the potentials in the theory \rf{sn} behave as
\be\label{appnew}
V(\phi) = {m^{2}\over 4} \left(1- {n(1+n)\over 1+2n}\, e^{-\sqrt 6 |\phi|}\right) \ ,
\ee 
up to higher order corrections in $e^{-\sqrt 6 |\phi|}$. These potentials look very different from each other at small $\phi$, but they have the same asymptotic behavior at large $\phi$, up to the coefficient in front of $e^{-\sqrt 6 |\phi|}$. Just like in \cite{Kallosh:2013hoa}, one can show that all of these models in the large $N$ limit have the same cosmological predictions \rf{alpha}, in good agreement with observations.
 
\begin{figure}[htb]
\begin{center}
\includegraphics[width=9cm]{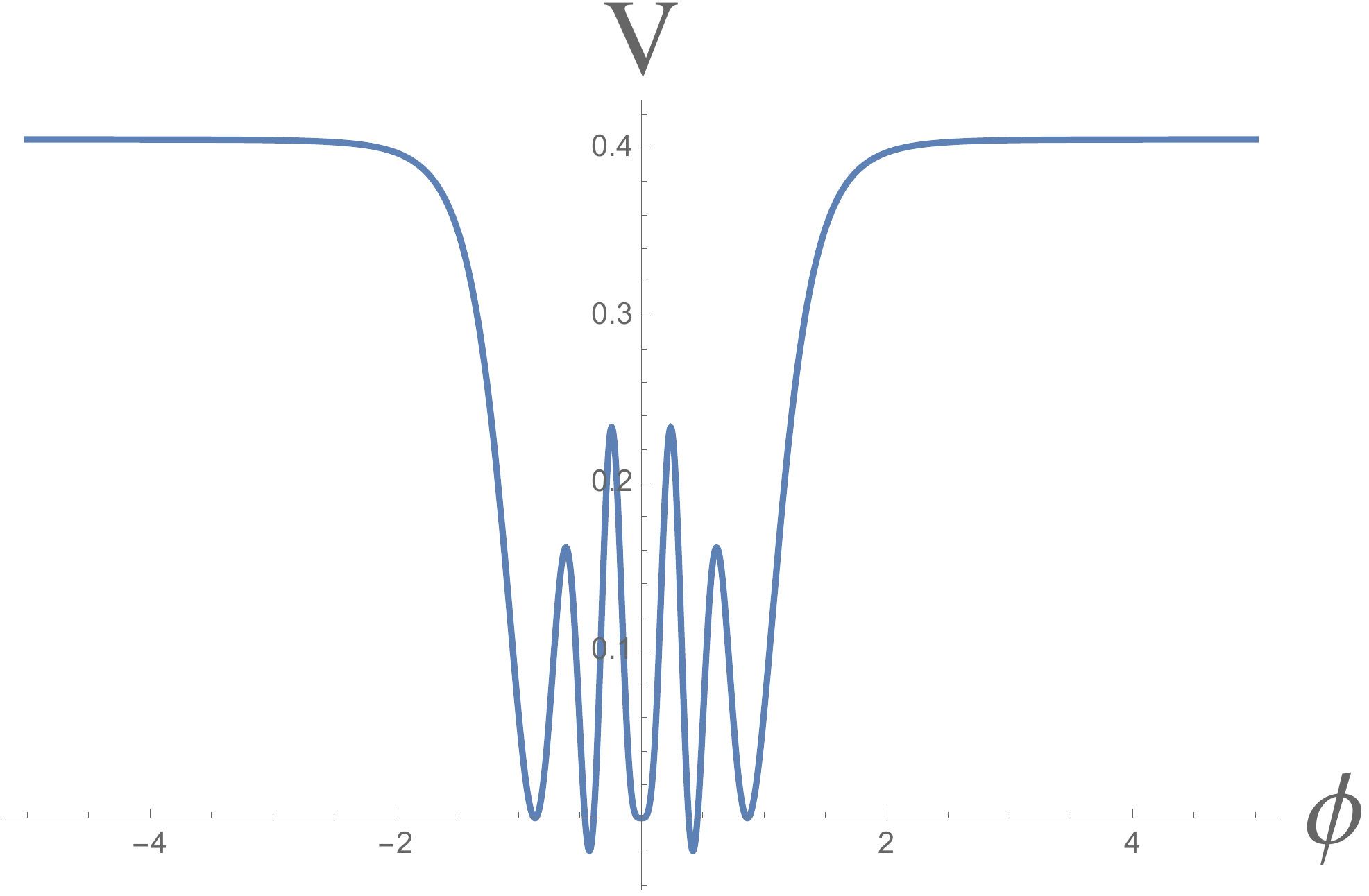}
\caption{\footnotesize The scalar potential in the model \rf{attr2}.  After inflation, the field rolls down to one of the two stable supersymmetric Minkowski vacua with $\phi \approx \pm 0.865$ and stays there.}\label{4}
\end{center}
\vspace{-0.5cm}
\end{figure}

As we already mentioned, one can consider  models with a more general superpotential  \cite{KL},
\be\label{attr}
W = W_{0}\sinh{\sqrt{3}\Phi}\, f(\tanh{\sqrt{3}\Phi}) \ ,
\ee
similarly to what happens in the theory of superconformal attractors \cite{Kallosh:2013hoa}. For a general choice of functions $f$, one obtains potentials which may behave very differently at small $\phi$, but which typically have a similar plateau behavior, asymptotically approaching a plateau as $V_{\infty}- c\, e^{-\sqrt 6 |\phi|}$, where $c$ is some constant. In those models with $V_{\infty} >0$, $c > 0$, where the field eventually rolls down to the near-Minkowski vacuum with a tiny cosmological constant, the observational predictions in the large $N$ limit approach the same attractor values \rf{alpha}. To give a particular example, one may consider a superpotential
\be\label{attr2}
W = W_{0}\sinh{\sqrt{3}\Phi}\, \sin^{2}(4\tanh{\sqrt{3}\Phi}) \ .
\ee
The inflaton potential in this theory is shown in Fig. \ref{4}. It has three Minkowski vacua and two AdS vacua. After inflation, the inflaton field rolls to one of the two absolutely stable supersymmetric Minkowski vacua with $\phi \approx \pm 0.865$. The observational predictions of this theory coincide with the predictions of the original GL model \rf{alpha}.

\section{Uplifting and SUSY breaking by adding a nilpotent field} 

The GL model and its generalizations considered above describe a potential with a Minkowski minimum with $V(0) = 0$, and with unbroken supersymmetry. As a next step towards a fully realistic model, it would be nice to describe our universe with a tiny positive cosmological constant, $V(0) \sim 10^{{-120}}$, and with a controllable level of supersymmetry breaking. Perhaps the simplest way to do it is to add to the theory a nilpotent chiral superfield $S$, following \cite{Ferrara:2014kva,Kallosh:2014via,Dall'Agata:2014oka,Kallosh:2014hxa}. Indeed, let us consider the model of two fields, the inflaton field $\Phi$ and a nilpotent field $S$ with the \K\ potential 
\be\label{shiftT}
K = -{1\over 2} (\Phi-\bar\Phi)^{2} +S\bar S
\ee
and the superpotential
\be\label{sT}
W = {m\over 6} \sinh{\sqrt{3}\Phi}\, \tanh{\sqrt{3}\Phi} + c S +d\ .
\ee
In accordance with the procedure outlined in \cite{Ferrara:2014kva}, one should calculate the potential using the standard methods, and then put $S = 0$ in the final expressions. In this case one can show that the potential of the inflaton field still has a minimum at $\phi = \chi = 0$, but the value of the potential at the minimum is given by
\be
V(0) =  c^{2}-3d^{2} \ .
\ee
For $c = \sqrt 3\, d$ one has a Minkowski minimum, as before. However if there is a slight mismatch between these two parameters, similar to what is supposed to happen in the string theory landscape  \cite{Eternal}, one can obtain any desirable value of the cosmological constant, including $V(0) \sim 10^{{-120}}$, and then one can use the anthropic considerations \cite{Linde:1984ir,Weinberg:1987dv} to explain why it should be small.

For $V(0) \sim 10^{{-120}}$, one still has $c = \sqrt 3\, d$ with enormous accuracy, but the values of $c$ and $d$ separately can be large. However, for  $c \gtrsim 10^{{-4}} m$ the shape of the potential at large $\phi$ becomes significantly distorted, so to preserve the cosmological predictions of the GL model one should take $c \ll 10^{-4}m \sim 7\times 10^{{-10}}$. 

The fermion masses can be computed in a particularly simple way if the conditions formulated in \cite{Kallosh:2014via,Dall'Agata:2014oka,Kallosh:2014hxa} are satisfied: $D_{S}W = c \not = 0$, and $D_{\Phi}W = 0$ at the minimum of the potential. In our model both of these conditions are satisfied. 
The inflatino mass at the minimum of the potential in this model is given by $m-m_{3/2}$, and the gravitino mass is given by
\be
m_{3/2} = d \approx  c/\sqrt 3 \ .
\ee
By a proper choice of the constant $c$, one can obtain supersymmetry breaking with the gravitino mass in a broad range, from nearly zero up to about $10^{-10}$ in Planck units.

\section{Discussion}

It is very difficult to obtain a consistent inflationary scenario in supergravity models with a single chiral superfield. 
 That is why the subsequent development in this area shifted towards development of models describing several superfields. It took 17 years since the development of the chaotic inflation scenario \cite{Linde:1983gd} until the simple chaotic inflation with a quadratic potential ${m^{2}\over 2}\phi^{2}$ was obtained in the models with two superfields and a shift-symmetric \K\ potential of the type of \rf{shiftT} \cite{Kawasaki:2000yn}. It took another 10 years until we learned how to find inflationary potentials of any desirable shape in this context \cite{Kallosh:2010xz}. The new generation of models is so flexible that it can fit any observational results for $n_{s}$ and $r$ \cite{Linde:2014nna}. The possibility to have inflation in the models with a single chiral superfield  was revisited only recently \cite{Ketov:2014qha,Linde:2014ela}, but it required a rather unusual choice of the \K\ potential.

That is why it is especially interesting that the very first version of the chaotic inflation in supergravity proposed more than 30 years ago in the context of a model of a single chiral superfield \cite{Goncharov:1983mw} did not require any modifications. This model in its original form, as well as its generalizations described in this paper, provides a very good fit to all presently available observational data.

\section*{Acknowledgments}

I am grateful to Alexander Goncharov and Renata Kallosh for inspiring discussions. My research is supported by the SITP, by the NSF Grant PHY-1316699, and by the Templeton Foundation grant `Inflation, the Multiverse, and Holography.'

\end{document}